\documentclass[lettersize,journal]{IEEEtran}
\usepackage{amsmath,amsfonts}
\usepackage{algorithmic}
\usepackage{algorithm}
\usepackage{array}
\usepackage{textcomp}
\usepackage{stfloats}
\usepackage{url}
\usepackage{verbatim}
\usepackage{graphicx}
\usepackage{cite}
\usepackage[dvipsnames,table]{xcolor}
\usepackage[hidelinks]{hyperref}
\usepackage[center]{caption}
\usepackage{flushend} 
\usepackage{fontawesome}
\usepackage{pifont}
\usepackage{bbding}
\usepackage{tikz}
\usepackage{multirow}
\usepackage{threeparttable}
\usepackage{subcaption}
\usepackage{lipsum}
\usepackage{caption}

\newcommand{\mypara}[1]{\vspace{2pt}\noindent{\textbf{#1}}}
\newcolumntype{P}[1]{>{\centering\arraybackslash}p{#1}}
\newcolumntype{M}[1]{>{\centering\arraybackslash}m{#1}}

\definecolor{myblue}{RGB}{22, 148, 178}
\definecolor{mygray}{RGB}{77, 77, 77}
\definecolor{tablegray}{RGB}{210, 210, 210}
\definecolor{tableblue}{RGB}{177, 221, 240}
\definecolor{mylilac}{RGB}{211, 223, 243}
\definecolor{myorange}{HTML}{FAD7AC}
\definecolor{mypink}{HTML}{E000E0}
\definecolor{mygreen}{HTML}{4AA024}

\begin{document}

\pagenumbering{gobble} 

\title{RISC-V Needs Secure “Wheels”: \\ the MCU Initiator-Side Perspective}
\author{Sandro Pinto, José Martins, Manuel Rodríguez, Luís Cunha, \\ Georg Schmalz*, Uwe Moslehner*, Kai Dieffenbach*, and Thomas Roecker*\\ OSYX Technnologies, Universidade do Minho, *Infineon AG\vspace{-0.5cm}
}



\maketitle


\begin{abstract}
The automotive industry is experiencing a massive paradigm shift. Cars are becoming increasingly autonomous, connected, and computerized. Modern electrical/electronic (E/E) architectures are pushing for an unforeseen functionality integration density, resulting in physically separate Electronic Control Units (ECUs) becoming virtualized and mapped to logical partitions within a single physical microcontroller (MCU). While functional safety (FuSa) has been pivotal for vehicle certification for decades, the increasing connectivity and advances have opened the door for a number of car hacks and attacks. This development drives (cyber-)security requirements in cars, and has paved the way for the release of the new security certification standard ISO21434. RISC-V has great potential to transform automotive computing systems, but we argue that current ISA / extensions are not ready yet.
This paper provides our critical perspective on the existing RISC-V limitations, particularly on the upcoming WorldGuard technology, to address virtualized MCU requirements in line with foreseen automotive applications and ISO21434 directives. We then present our proposal for the required ISA extensions to address such limitations, mainly targeting initiator-side protection. Finally, we explain our roadmap towards a full open-source proof-of-concept (PoC), which includes extending QEMU, an open-source RISC-V core, and building a complete software stack.

\end{abstract}


\section{Introduction}


The automotive industry is currently experiencing a massive transformation, with cars being currently coined as high-performance computers on wheels. First, traditional electrical/electronic (E/E) architectures, where functionalities were typically scattered across tens/hundreds of single-purpose Electronic Control Units (ECUs), are shifting toward centralized Domain and Zone architectures \cite{Staron2021, mauser2024}. At the same time, with the rise of the vehicle-to-everything (V2X) concept, cars are also becoming increasingly connected, and thus, security has been promoted as a first-class citizen in the automotive list of requirements \cite{unece2023grva}.

While functional safety (FuSa) standards, e.g., ISO26262, have been pivotal for the certification of vehicles (and respective systems and components) for decades \cite{palin2011iso}, automotive security approaches have often been very specific to respective applications, with OEMs, Tier 1, and Tier 2 suppliers formalizing their own approaches to address cybersecurity (e.g., tuning protection). Nevertheless, over the last few years, and mainly steamed by this increasing need for (cyber-)security, the automotive industry has been particularly focused on devising a more standardized framework to protect vehicles and tackle cybersecurity for automotive products, which resulted in the new international standard coined ISO21434 \cite{iso21434, Ebrahimi2023}.

While the standard remains mostly abstract regarding conceptual approaches (i.e., the standard dictates the structuring of development, risk assessment, and scaled methodological requirements), the genericity of security requirements mandates the introduction of additional features to the MCU. This development is essentially driven by the novel E/E-architectures, with the explicit goal of reducing the number of physically separate ECUs \cite{Burgio2016, Staron2021}. Consequently, integration density per ECU is growing, resulting in multiple MCUs becoming virtualized (vMCU), i.e., mapping multiple logical partitions within a single physical MCU (pMCU) \cite{Martins2020, Pinto2019, Oliveira2024}. Typically, vMCUs are of different integrity levels (either related to their applicative nature/goal or due to multi-party development), thus requiring isolation between partitions to avoid requalification effort (so called freedom-from-interference, or simply FFI) \cite{Kloda2019, Bechtel2019, Martins2023}. FFI relies on hardware-based mechanisms allowing configurable restriction of access rights, which is particularly important if physical resources (e.g., central memory) are shared among vMCUs. We term the underlying hardware concept ‘Isolation Mechanism’ (IM). 

Well-established ISAs and vendors in the MCU spectrum have been proposing a set of IMs to address modern safety and security requirements for the Internet of Things (IoT) and automotive. These IM can basically be grouped into two major classes of primitives: (i) the ones implemented at the CPU level and (ii) the ones implemented at the system level (aiming at enforcing access control in mixed-criticality environments and for non-CPU initiators such as DMA). While the former are mostly implemented through standard ISA memory protection facilities (or technologies), the latter are typically implemented through custom vendor infrastructures. For example, Arm's Cortex-M line is equipped with a CPU-level Memory Protection Unit (MPU) \cite{Michele2022, Xi2024}, while the newer Armv8-M further endowed Cortex-M MCUs with TrustZone (i.e., additional CPU-level controllers like the Security Attribution Unit - SAU) \cite{TZSurvey2019, Xi2024}. At the same time, and mostly driven by the real-time automotive requirements, Arm has been pushing for the Armv8-R architecture, which includes already hardware virtualization support via dual-stage MPU \cite{armv8r_whitepaper}. 
However, we note that system-level protection is employing heterogenous proprietary solutions (e.g., Implementation-defined attribution unit, system-level MPUs or other mechanisms for rule enforcement). 
While all technologies provide security through isolation, particularly compatibility and scalability are vastly different and still a topic of debate in industry and academia (e.g., in context of Software-defined vehicle). In this context, the number of distinguishable partitions (referred as IDs) plays an important role, as it has notable impact to area.

RISC-V is seen as a game-changer in the automotive industry. Many “voices” have been very vocal about the pivotal opportunity for RISC-V to transform automotive computing systems in the years to come. While some RISC-V vendors have been pushing and marketing the (imminent) availability of certified-ready RISC-V MCUs for multiple Automotive Safety Integrity Levels (e.g., ASIL-B, ASIL-D), very little has been discussed and done for compliance with the automotive security standard ISO21434. While it is true that, from an ISA perspective, RISC-V International is currently specifying key hardware isolation primitives or IMs to enable virtualized MCU functions, e.g., the S-mode PMP (and SPMP for hypervisor) \cite{riscv_spmp_2023, riscv-spmp-hypervisor} and WorldGuard (WG) \cite{wg_spec} are being currently ratified and/or making its way for a “fast-track” ratification, we argue that, fundamentally, all these technologies (and mostly WorldGuard) are also not yet ready to address the requirements of next-generation automotive applications and requirements.

In this paper, we start by providing our critical perspective on the existing limitations of WorldGuard in interoperating with other key extensions to enable virtualized MCU functions and in scaling per the needs of foreseen automotive applications and ISO21434 directives. Then, we present our proposal for extending the WorldGuard specification to (i) interoperate with the Hypervisor extension (in light of the ongoing SPMP for Hypervisor) and (ii) increase up to 128 IDs/worlds. We focus our proposal on the initiator-side mechanisms required to realize the respective ISA-extensions and defer any further developments on the resource-side for future work. We conclude the paper by laying out the roadmap towards a full open-source Proof-of-Concept (PoC), which includes extending the WG QEMU, implementing WG (and extensions) in an open-source RISC-V CPU, and building a software stack based on the Bao hypervisor \cite{Martins2020}.

In summary, with this work, we make the following contributions:
\begin{enumerate}

  \item We analyze modern E/E architectures requirements and we discuss the major RISC-V ISA gaps for materializing secure real-time RISC-V MCUs for automotive;
  
  \item We propose a set of extensions to the SPMP (for hypervisor) and WorldGuard specifications to address and mitigate the identified limitations;

  \item Finally, we discuss the ongoing activities and roadmap to materialize a full open-source hardware-software Proof-of-Concept (PoC) of the proposed extensions and system.
  
\end{enumerate}

\section{Background}

\subsection{RISC-V Privileged ISA}

RISC-V distinguishes itself from traditional platforms by offering a free and open ISA featuring a modular and highly customizable extension scheme. The RISC-V architecture \cite{privileged_isa} divides its execution model into 3 privilege levels: (i) machine mode (M-mode), i.e., the most privileged level, hosting the firmware which implements the supervisor binary interface (SBI) (e.g., OpenSBI); (ii) supervisor mode (S-Mode) for running General-Purpose OS (GPOS), e.g., Linux, that require virtual memory management \footnote{We mention Linux as the typical example of a guest OS, although this paper addresses MMU-less systems where Linux would not be able to run.}; (iii) user mode (U-Mode) for running user-level applications. The modularity offered by the RISC-V ISA seamlessly allows for implementations at distinct design points, ranging from small embedded platforms with just M-mode support to fully blown server class systems with M/S/U.

\mypara{RISC-V Hypervisor Extension.} The RISC-V privileged architecture specification introduced hardware support for virtualization through the Hypervisor extension \cite{Sa2022, Sa2023}. This extension execution model follows an orthogonal design where the supervisor mode (S-mode) is modified to a hypervisor-extended supervisor mode (HS-mode). There are two new privileged modes, i.e., virtual supervisor mode (VS-mode) and virtual user mode (VU-mode). The Hypervisor extension also defines a second stage of translation (G-stage) to virtualize the guest memory by translating guest-physical addresses (GPA) into host-physical addresses (HPA). The HS-mode operates like S-mode but with additional hypervisor registers and instructions to control the VM execution and G-stage translation. Note that, in the scope of this paper, this second stage of translation (i.e., second-stage MMU) is replace by a second-stage PMP.

\subsection{SPMP in a nutshell}

\begin{figure}[t!]
    \centering
    \vspace{-0.5cm}
    \begin{subfigure}[b]{0.4\linewidth}
        \centering
        \includegraphics[height=7cm]{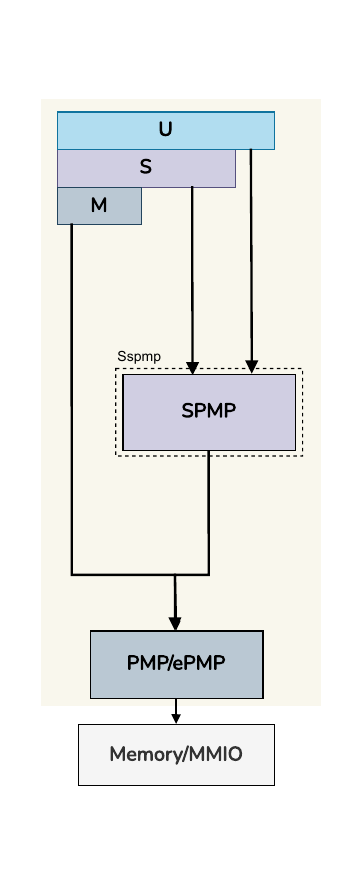}
        \vspace{-0.5cm}
        \caption{Base SPMP.}
        \label{fig:spmp-base}
    \end{subfigure}
    \hfill
    \begin{subfigure}[b]{0.5\linewidth}
        \centering
        \vspace{-1cm}
        \includegraphics[height=7cm]{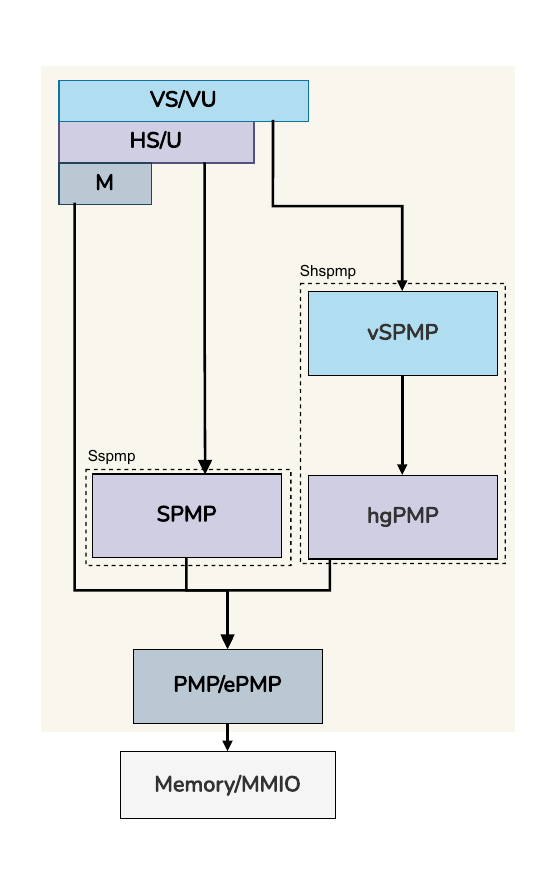}
        \vspace{-0.5cm}
        \caption{SPMP for Hypervisor.}
        \label{fig:spmp-hyprvisor}
    \end{subfigure}
    \caption{SPMP with and without the Hypervisor Extension.}
    \label{fig:spmp-base-models}
\end{figure}

\label{sec:spmp-nutshell}
In the RISC-V lingo, the Memory Protection Unit (MPU) is refered to as Physical Memory Protection (PMP). Initially, only the highest privilege mode, i.e., Machine-mode (M-mode), featured a PMP, to enable isolation of firmware and OS-managed resources to be used as a primitive for static resource partitioning and Trusted Execution Enviroment (TEE) creation. Due to the need to improve latency, determinism, and safety (over GPOS), Real-Time operating systems (RTOS) are typically relying on OS-managed memory protection primitives/facilities to enforce user-level application isolation. Consequently, the RISC-V community proposed the Sspmp extension \cite{riscv_spmp_2023} that replaces the Memory Management Unit (MMU) by a S-Mode PMP, thus SPMP, checking accesses from both User and Supervisor modes (see Figure \ref{fig:spmp-base}). The SPMP follows essentially the same design as the (e)PMP, featuring an implementation-defined number of entries and a mode bit that differentiates between Supervisor and User access permissions. On each context switch, the OS must swap user entries. Depending on the number of regions, this might be an expensive operation; thus, the SPMP introduced a context-switch optimization mechanism unavailable on the original PMP. This mechanism defines a set of registers that enable/disable (bit-wise) SPMP entries. Thus, if enough entries are available, a context switch might resume to a write to these registers, effectively disabling entries related to suspended tasks and enabling entries associated with newly scheduled tasks. As of this writing, the SPMP extension is not yet ratified; however, it has already been submitted for prior consideration of the Architecture Review Committee (ARC). The next step is the public review and then ratification.   


\mypara{SPMP for hypervisor.} The SPMP is currently being extended to support the Hypervisor extension by defining a dual-stage PMP \cite{riscv-spmp-hypervisor} as depicted in Figure \ref{fig:spmp-hyprvisor}. The first stage, dubbed the virtual SPMP or vSPMP, is controlled by the VMs running in VS-mode, while the second stage is controlled by the hypervisor, effectively enforcing isolation among VMs. The existing proposal features two separate PMPs controlled by HS mode, where the (i) first (the baseline SPMP) mediates accesses from HS/U, and the (ii) second, dubbed hgPMP, mediates access from VMs, i.e., from VS/VU modes. 

\subsection{WorldGuard in a nutshell}

While the RISC-V (S)PMP provides isolation by enforcing access control through a set of well-defined memory configuration rules (PMP entries) on physical memory accesses initiated at the hart level, the RISC-V WorldGuard \cite{wg_spec} provides isolation in a complete holistic fashion (i.e., system-wide approach) through the concept of \textit{Worlds}. Worlds are simply system execution contexts that include both agents (i.e., system components that initiate transactions) and resources (i.e., system components that respond to transactions). Per the WG specification, Worlds are identified by a hardware \textit{World Identifier} (WID). The maximum number of unique WIDs is platform-specific (well-defined by \texttt{NWorlds}) but capped to a maximum of 32 Worlds\footnote{While the WG specification claims that 2-8 WIDs are sufficient for many use cases, we demonstrate in Section III that this is not necessarely the case for automotive applications in Zone/Domain architectures}. WG aims to provide static allocation of agents and resources to Worlds (typically configured by M-mode firmware or a TEE at boot time), and any efficient implementation of dynamic changes is outside of the scope of the current specification. In contrast to the (S)PMP, where permission checks are performed directly at the hart, WG does not define any specific method for propagating and checking the WID on the platform / bus, i.e., it is completely platform-specific. Thus, different platforms may implement different checking schemes, with bus fabrics supporting WIDs in implementation-specific ways.

\mypara{RISC-V ISA WG Extensions.} WG specifies three different levels of support at the hart level, by providing three different ISA extensions (see Table \ref{tab:wg-isa-ext}). For the first level, which fixes (hard-wires) the WID for all privilege modes on a hart, there is no ISA extension. For the second level, which enables M-mode to control the WID of lower-privilege modes, WG requires the Smwg extension. Finally, the third level introduces the Smwgd extension, which further enables M-mode to delegate to (H)S-mode the ability to assign the WID of lower-privilege modes, thus adding the Sswg extension to (H)S-mode.

\begin{table}[t]
\caption{WorldGuard ISA Extensions CSRs}
\center
\begin{tabular}{|c|l|p{5cm}|}
\hline
\centering\textbf{Extension} 
& \textbf{CSRs}
& \textbf{Description} 
\\ 
\hline\hline
\multirow{1}{*}{Smwg} 
 & mlwid   &  Defines the WID used for all lower privilege mode accesses in the current hart. \\ \cline{1-3}   
\multirow{1}{*}{Smwgd} 
 & mwiddeleg   &  Bit vector setting the WIDs delegated to S/HS mode.  \\ \cline{1-3}
\multirow{1}{*}{Sswg} 
 & slwid  &  Defines the WID used for U, VS and VU mode accesses in the current hart.   \\ \cline{1-3}   \hline
\end{tabular}
\label{tab:wg-isa-ext}
\end{table}

\section{Use Cases, Motivation, and RISC-V Gaps}

\subsection{Real-time Automotive Isolation: Use Cases}

In the field of automotive real-time applications, IM has to fulfill further requirements, i.e., beyond the pure isolation capabilities: (i) limited impact on access latencies within the system; (ii) flexible/application-oriented configuration model; (iii) absence of systematic-faults up to ASIL-D; and (iv) random-event robustness. Furthermore, if the MCU offers hardware virtualization support, IM has to address isolation among virtual CPUs / machines / harts.

Typical isolation use-cases are: (i) integration of tactic and strategic functions into a single MCU ("tactic" refers to functions having direct impact on vehicles motion, whereas strategic functions address a higher abstraction layer, e.g., forecasting without direct control); (ii) routing of multi-purpose data-streams with different criticality targets (e.g., charging station and end-point ECU); and (iii) virtualized security-island, i.e., either replacing or complementing a classical hardware security module (HSM). It is worth noting that these use-cases are neither mutually exclusive, nor complete, as for defining E/E-architectures there are still various models in evaluation by OEM/TIER. In addition, the nature of value-chained multi-party business mandates for strict isolation already for liability reasons.

\subsection{Real-time Automotive Isolation: Motivation}


In order to assess the number of isolated partitions to be formed within a single MCU implementing multiple vMCUs, we base our analysis in two major drivers:

\begin{itemize}

  \item \textbf{Inter-vMCU.} Each vMCU (which was formerly mapped to a dedicated physical computing unit) needs isolation from all others deployed to the same physical MCU. Such separation is of static nature, as the vertical integration either reuses former / legacy stacks (with known resource footprint) or such assignment is performed during system design-time.

  \item \textbf{Intra-vMCU.} Applications running in a vMCU with different criticality shall be separated, i.e., systems with mixed ASIL (or CAL, SIL, etc.) or of different source (e.g., third-party SW). This strategy enables cost-efficiency and is foundational for reliable value-chaining.

\end{itemize}

It should be noted that OS-scheduled cohesive units, i.e., tasks or ISRs contributing to the same functional goal (e.g., classified in trusted and non-trusted), are typically isolated via built-in OS-level memory protection mechanisms (e.g., RTOS leveraging MPU/PMP facilities). Nevertheless, in general, this can neither replace Inter- nor Intra-vMCU ID-based separation, considering the following arguments:

\begin{itemize}

  \item \textbf{Inter-vMCU.} In general, legacy stacks running on formerly physically separated computing units do not contain memory protection (MPU/PMP) programming routines. Thus, applying separation rules can be highly restrictive. Furthermore, modification of legacy stacks is unwanted to minimize software re-qualification efforts.

      \item \textbf{Intra-vMCU.} Target of integration are single- or multi-core stacks of different criticality. Due to cost-reasons, each of these stacks is typically only developed to a specific integrity level, sufficient to fulfill the respective application needs (e.g., ASIL-D for drive-train related control and QM for commodity/infotainment). In the absence of system-level protection, PMP-programming by the stack with lowest integrity would determine the vMCUs final integrity, which thwarts the idea of mixed-criticality integration.

\end{itemize}

It should be noted that integration of high-privilege/-integrity firmware (FW) in M-Mode for ePMP-programming could be at first hand regarded as an alternative; however, this has the following caveats. First, we argue that in most implementations there would be limitations related to the required number of memory protection regions (i.e., PMP entries), in particular considering the existence of non-consecutive I/O-ranges (e.g., routing use-cases). Second, we argue that such approach adds latency to the VM-switch for reprogramming PMP entries (rather than only changing the WID-setting), which is highly undesirable when one/a few of the integrated vMCUs implement real-time characteristics. Finally, we argue that PMP primitives do not offer protection for non-CPU initiators, e.g., DMAs and crypto accelerators, which are widely used in (automotive) MCUs.

Based on the articulated rationale, we suggest to estimate upper thresholds for the required isolated partitions based on the following ground-rules:

\begin{itemize}

    \item \textbf{One ID per non-CPU initiator}. This shall allow resolution down to respective atomic sub-functionality (e.g., for DMA one ID per channel), which we call atomic non-core master-function (ANM). These are typically scaling with the number of CPUs.

    \item \textbf{One ID per physical/non-virtualized CPU/Hart and privilege level (S, U)}. 
    
    \item \textbf{One ID per virtual CPU/Hart and virtualized privilege level (HS, VS, VU)}. Inclusion of HS-Mode in this scheme allows separation in case core-local hypervisors are used on a single MCU.
    
    \item \textbf{One ID is per physical Hart to accommodate M-Level separation}. From an architectural perspective, M-Mode SW/FW is considered to be of highest integrity-level, handling fundamental basic tasks (Start-up, Delegation, SysCalls, etc.). This would imply that a single ID for the whole system could be sufficient to identify M-Level SW. However, this would not consider that M-Level SW-components deployed to different Harts could come from different vendors (and thus requiring separation). We assume this to be the case till automotive SW-stacks have been adapted to RISC-V at scale.
    
    \item \textbf{One ID per secundary/auxiliar Hart.} Typically, SoCs integrate smaller cores (M,U) next to main cores. The former are used for house-keeping, security, power management, etc. In such configurations, the respective PMPs are assumed to implement only small number of regions and consequently distinct IDs for different-criticality data  accessed in U-mode might need to be reserved. In such cases, we foresee for small cores an additional ID, particularly for small-medium SoC configurations. We do not see this as being applicable for main-cores, as these are considered to provide sufficient PMP-capacity.

We note that a different model would need to draw assumptions on number of independent SW-partitions. Consequently, such approach would be application-dependent and insufficient to determine upper thresholds.

\end{itemize}

\begin{table}[!t]
\caption{MCU properties based on different assumed system configurations and estimated number of WIDs.}
\label{tab:table_wgid}
\resizebox{0.5\textwidth}{!}{%
\centering
\begin{threeparttable}
\scriptsize
\begin{tabular}{|m{1.7cm}|M{1.0cm}|M{3.6cm}|M{0.7cm}||M{0.7cm}|}
\hline

\centering\textbf{System Config.} 
& \textbf{Number of Harts}
& \textbf{Privilege Levels} 
& \textbf{ANMs} 
& \textbf{WIDs} 
\\ 
\hline\hline

\textit{small}
& 2
& 2 (M+U)
& 10
& \textbf{16}
\\ 
\hline

\textit{medium}
& 4
& 3 (M+S+U)
& 30
& \textbf{43}
\\ 
\hline

\textit{high}
& 6 + 2
& 3 (M+HS+VS+VU) + 3 (M+S+U) + 1 (M+U) + 1 (M+S+U)
& 50
& \textbf{82}
\\ 
\hline

\end{tabular}
\end{threeparttable}%
}
\end{table}

To provide the resulting estimations in terms of required WIDs, we consider three MCU configurations which we refer to as \textit{'small'}, \textit{'medium'}, and \textit{'high'}, summarized in Table \ref{tab:table_wgid}. The \textit{small} configuration implements two Harts with two privilege levels (M+U) and 10 ANMs. The \textit{medium} configuration includes four Harts with three privilege levels (M+S+U) cores and 30 ANMs. 
These configurations refer to ECUs for end-point and small-mid zone applications, respectively.
The \textit{high} configuration is distinctly different from \textit{small} / \textit{medium} by providing support for virtualization and assigning smaller auxiliary Harts for house-keeping. Thus, we require here an assumption on the number of deployed VMs and the number of auxiliary Harts. As the former is naturally limited by the VM-scheduling overheads and induced latency, we assume here exemplary a half-virtualized case: 3/6 main-Harts host two VMs each, while the other 3/6 run non-virtualized. In addition, we assume one auxiliary Hart with two privilege levels (M+U), another Hart with three privilege levels (M+S+U), and 50 ANMs.

Based on the described rationale, we calculate / estimate 16, 43, 82 WIDs required for the \textit{small}, \textit{medium}, \textit{high} system configurations, respectively. We argue that these examples provide enough evidence that the current upper threshold on the maximun number of 32 WIDs, per the WG specification, is a too strict limitation for mid- to large Zone/Cross-domain Controllers.

\begin{figure}[!t]
\centering
\includegraphics[width=\linewidth]{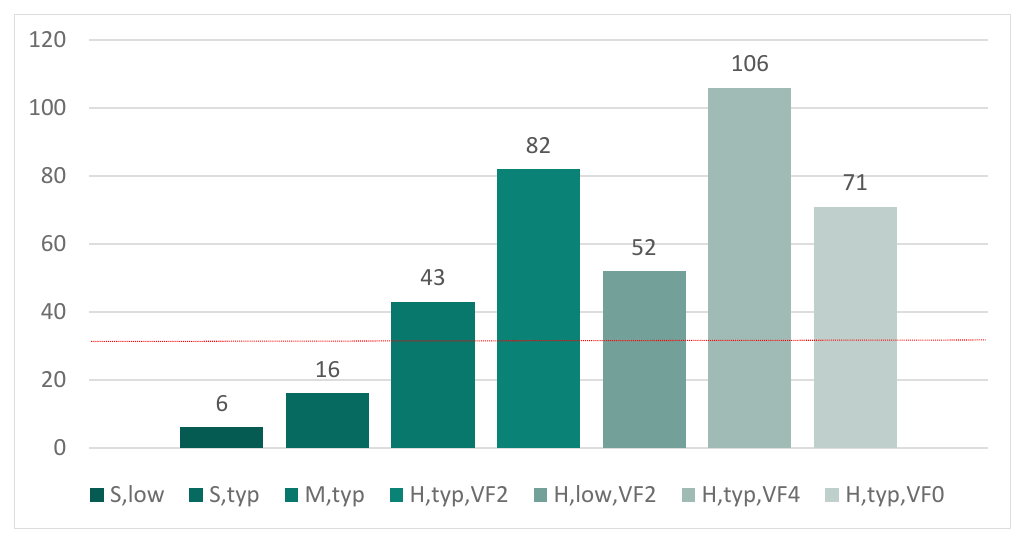}
\caption{Estimations for the number of WIDs required for the example configurations with varying number of harts and privilege levels.}
\label{fig_wg-worlds}
\end{figure}

To complement the WIDs estimation and provide stronger evidence, we augmented the number of configurations in addition to the \textit{small} (S,typical $\rightarrow$ 16), \textit{medium} (M,typical $\rightarrow$ 43), and high (H,typical,VF2 $\rightarrow$ 82). We modified the number of ANM (low,typical) for \textit{small} and \textit{medium} devices. For the \textit{high} configuration, we also modified the virtualization profile (number of virtualized cores and number of VMs per Hart $\rightarrow$ VF). Additionally, (S, low) has been included as lower-end reference configuration (2 M+U Harts and 2 ANMs, e.g., 1 DMA-channel per Hart, no additional U-mode ID), which is referenced in the WorldGuard specification as being sufficient for many use cases. Figure \ref{fig_wg-worlds} presents the estimated WIDs required for such configurations.

We observe that strongly virtualized systems (H,typical,VF4 $\rightarrow$ 5/6 main Harts virtualized with 4 VMs each) drive the number of potentially distinct partitions even further, in particularly when comparing to a high-end non-virtualized system (H,typical,VF0). Nevertheless, we note that even in high-end Zone-Controllers such a configuration would most likely mark an upper threshold due to overhead (latency, power-consumption) induced by virtualization and limited number of localized functions (assuming six zones in the high-end segment). Comparing H,typical,VF2 and H,low,VF2, we would like to highlight that in case of high number of ANMs, a initiator-side mechanism for the latter might be an alternative in order to keep the number of IDs (impacting checker-structures) below 64, which might provide an efficient architectural alternative. Nevertheless, we defer such discussions for future work.

\subsection{Real-time Automotive Isolation: RISC-V ISA Gaps} \label{sec:riscv-gaps}

While WorldGuard itself offers a lean, easy model for initiator-side identification on privilege-level granularity, based on the arguments justified in the previous subsection, we strongly believe that two aspects are missing in the current form of the specification:

\begin{itemize}

  \item Support for virtualized environments, i.e., WID based identification of additional privilege-levels introduced via the Hypervisor extension. This aspect is especially important for bare-metal application w/o address-translation (so called ‘real-time virtualization’), where the Hypervisor plays an essential role in the safety-/security-concept. Though certain reference is made in the original specification, it maps the same ID to both VS/VU-levels which is insufficient, if distinction between OS- and User-Task in virtualized environments are required.

  \item Support for more than 32 WIDs, i.e., larger number of distinguishable partitions. As fundamented in the previous subsection, this need is driven by growing MCU-complexity, integrating more and more (at least partially virtualizable) initiators and cross-combinations of different criticality/integrity (e.g., ASIL x CAL).

\end{itemize}

In addition, we argue that the current specification of the SPMP for Hypervisor with separate SPMPs for non-virtualized and virtualized modes is inefficient. This mainly stems from the fact that, because the \textit{a priori} static partition of costly SPMP entries between the baseline SPMP and the hgPMP, for some configurations, some entries might be left unused in one of these components while more entries are required for the other. Therefore we opt to migrate the design-time allocation of ranges to become a configuration-time option.


\section{RISC-V CPU Extension}

In this section, we propose extensions for the existing SPMP for hypervisor and WorldGuard specifications, addressing identified inefficiencies or limitations.

\subsection{Unified SPMP for Hypervisor}

\begin{figure}[!t]
    \centering
    \vspace{-0.5cm}
    \begin{subfigure}[b]{0.5\linewidth}
        \centering
        \includegraphics[height=7cm]{images/spmp-separate.pdf}
        \vspace{-0.5cm}
        \caption{Separate.}
        \label{fig:spmp-models-separate}
    \end{subfigure}
    \hfill
    \begin{subfigure}[b]{0.4\linewidth}
        \centering
        \vspace{-1cm}
        \includegraphics[height=7cm]{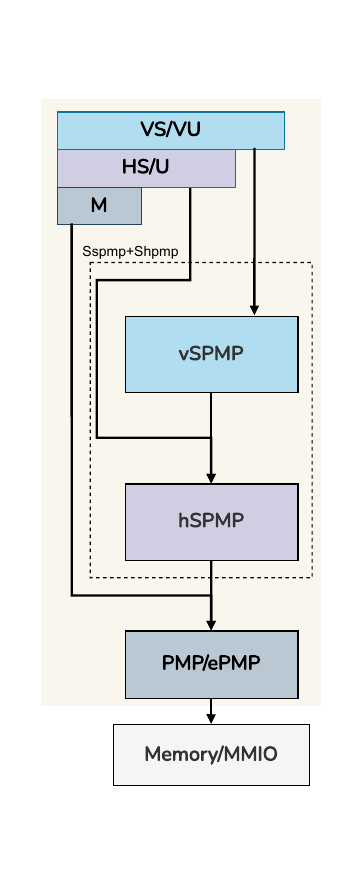}
        \vspace{-0.5cm}
        \caption{Unified.}
        \label{fig:spmp-models-unified}
    \end{subfigure}
    \caption{Hypervisor SPMP models.}
    \label{fig:spmp-moels}
\end{figure}

As explained in Section \ref{sec:spmp-nutshell}, the original SPMP specification for hypervisor, besides the vSPMP in control of VS mode, adds a separate second stage SPMP (hgPMP) in control of the HS-mode hypervisor, that checks guests accesses (see Figure \ref{fig:spmp-models-separate}). The hgPMP is completely separate of the original SPMP that checks HS- and U-mode accesses. The main hindrance of this model is that the costly SPMP entries are statically partitioned between the two SPMPs controlled by the hypervisor. In this sense, if, for example, the hypervisor does not use all entries in the SPMP, it cannot reuse them for the second stage hgPMP even if it requires more entries to setup the guest's address space.

In this work, we introduced the SPMP for Hypervisor unified model where only a single SPMP is controlled by the hypervisor. In this model, execution under either VS- or VU-mode mimics the execution of a user-mode application on the baseline SPMP, meaning the second stage SPMP is just the baseline SPMP (see Figure \ref{fig:spmp-models-unified}). In our point of view, besides being a much cleaner extension, this approach prevents the \textit{a priori} partition of SPMP entries between the hypervisor and the virtual modes, reducing the waste of possible unused entries if, for example, virtualization is not used. The other modification added by this model is the addition of the \textit{hspmpswitch} which replaces the function as the original \textit{spmpswitch} when virtualized modes are executing. The addition of this register helps the hypervisor clearly distinguish which entries are used for itself (HS/U) and which are used for the guest (VS/VU) in the unified hSPMP without incurring in merging hypervisor and guest entries using the original shared permissions encoding, greatly simplifying SPMP entry management code.

\subsection{WorldGuard CSRs extension}
\label{sec:wg-ext}

As described in Section \ref{sec:riscv-gaps}, WorldGuard is essentially comprised of initiator-side identification and resource-side MMIO-controlled checkers, where the former adds respective CSRs. For the moment, we focus our scope on the initiator-side, as we observe uncertainty and ongoing concurrent development between WorldGuard-proposed PMP-scheme based checkers and IOPMP specification activities.

In order to address the limitations of existing specifications (Section \ref{sec:riscv-gaps}), we propose two new extensions (listed in Table \ref{tab:h-feat}):

\begin{itemize}

  \item The \textbf{\textit{Shwgd} extension}, to add another set of WID-registers to allow independent configuration of identification for virtual-supervisor (VS)-level: \textit{hslwid} and \textit{hwiddeleg} which configure VS-ID and the ID-mask for lower privilege, respectively. Both are configurable from HS-mode. It should be noted that this implies that mlwid and mwiddeleg are addressing HS-level and lower. Furthermore, \textit{slwid} is renamed \textit{vslwid}, which allows configuration of VU-Level ID independently from VS-Level. When executing in VS-mode (i.e., Virtualization Mode V=1), \textit{vslwid} is accessed under the same address as original \textit{slwid}, and constraints from \textit{mwiddeleg} apply to \textit{hslwid} and lower privilege deleg-type registers. For V=0, \textit{hslwid} and \textit{hwiddeleg} have no effect, i.e., U-mode ID is determined by \textit{slwid}, itself constrained by \textit{mwiddeleg}.

  \item The \textbf{\textit{Slwgd} extension}, to augment the deleg-type registers to support up to 64/128 IDs, i.e., support for additional bitvectors is required. Therefore, we introduce registers \textit{mwiddelegh}, \textit{hwiddelegh} (+ corresponding registers with subscript h2, h3 in case 128 IDs are required for system-under-design). While for RV32, a two-/four-instruction sequence (using explicit encoded addressing in 12bit-space) is required, to consistently operate on the deleg-CSRs, a 64bit-implementation shall allow the same in a single-/double-instruction sequence. Note that this is similar to the approach applied for the User-Level counters.

\end{itemize}

As the above scheme adds up to 8 additional CSRs (\textit{hslwid}, \textit{hwiddeleg} for H-extension and \textit{mwiddelegh,h2,h3}/\textit{hwiddelegh,h2,h3} for up to 128WIDs), investigation is required to assess how registers can be efficiently mapped to address space. 'Efficiently' shall not be limited to minimization of used address space, but particularly consider the impact of (re)programming-induced latency, which we consider as equally important.

\begin{table}[t]
\caption{Proposed WorldGuard ISA Extensions CSRs for Hypervisor Support and WID extension.}
\center
\begin{tabular}{|c|l|p{4.5cm}|}
\hline
\centering\textbf{Extension} 
& \textbf{CSRs}
& \textbf{Description} 
\\ 
\hline\hline
\multirow{7}{*}{Shwgd} 
 & hslwid      &   Defines the WID used for all lower virtualized mode accesses in the current hart. \\ \cline{2-3}   
 & hwiddeleg   &  Bit vector setting the WIDs delegated to VS mode. \\ \cline{2-3}
 & vslwid      &  Defines the WID used for VU mode accesses in the current hart. \\ \cline{2-3}
 \hline
\multirow{12}{*}{Slwgd} 
 & mwiddelegh    &  Bit vector setting the WIDs 32-63 delegated to S/HS mode.  \\ \cline{2-3}
 & mwiddelegh2   &  Bit vector setting the WIDs 64-95 delegated to S/HS mode.  \\ \cline{2-3}
 & mwiddelegh3   &  Bit vector setting the WIDs 96-127 delegated to S/HS mode.  \\ \cline{2-3}
 & hwiddelegh    &  Bit vector setting the WIDs 32-63 delegated to VS mode.   \\ \cline{2-3}
 & hwiddelegh2   &  Bit vector setting the WIDs 64-95 delegated to VS mode.   \\ \cline{2-3}
 & hwiddelegh3   &  Bit vector setting the WIDs 96-127 delegated to VS mode.  \\ \cline{1-3}
\end{tabular}
\label{tab:h-feat}
\end{table}

\subsection{Programming Model}

We consider the following programming model: deleg-type registers are assumed to be (pseudo-)static (i.e., configured during start-up or infrequent macroscopic repartitioning of the system), while lwid-type registers are dynamically reconfigured (VM-Swap or OS-controlled change of application). That is, SPMP configurations are dynamically swapped based on fixed sets, known at compile-time.

Though beyond the scope of this paper, we would like sketch our assumptions on the resource-side checkers: (i) memory-related checkers are range-configurable, while this is not the case for peripherals; and (ii) their configurations (WID, ranges) are also (pseudo-)static, i.e., are not or very seldomly changed during run-time.
Furthermore, it assumes M-Level IDs are fixed and do not require further initialization.

\mypara{Initialization Flow.} During system initialization, firmware running in M-mode will start by deciding which memory regions/peripherals it exclusively requires for its own operation and consequently configure and lock the respective checkers. While M-mode has a distinct WID, firmware may decide to keep further IDs reserved, by not delegating them to lower privilege (hypervisor or OS). Then it will set \textit{mlwid} and \textit{mwiddeleg}, i.e., define the identification for hypervisor and potentially reserved IDs. When the hypervisor starts executing, the hypervisor will perform a similar sequence, by first deciding which WIDs will reserve for itself, and which WIDs it will delegate to each VM. Once it has finished its reservation sequence, it configures \textit{hslwid} with the VM to be scheduled first, and respective \textit{hwiddeleg}. Checkers which have not been configured/locked by hypervisor could be allocated incrementally during VM-runtime. 
When restoring the context of a given vCPU, the hypervisor will set \textit{hslwid} and \textit{hswiddeleg} according to the WIDs delegated to its VM. Then, it will set hSPMP entries with the memory regions assigned to that VM and entries to passthrough access to all MMIO regions given that access control to peripherals is already guaranteed by WG checkers. 
We consider this assumption hold true in a system with (quasi-)static partitioning, i.e., fixed allocation of certain peripheral-/memory-space to vMCU. 

\section{Roadmap and Next Steps}

\mypara{QEMU extension.} WorldGuard support for QEMU was recently released by SiFive \cite{wg-qemu}, but is not upstreamed yet (as WorldGuard specification still needs to go through ratification). We will extend the existing WorldGuard model to (i) include the proposed the CSRs extensions described in Section \ref{sec:wg-ext} as well as (ii) have interoperable support between WorldGuard and the RISC-V SPMP for hypervisor support (that we are currently also extending in QEMU).   

\mypara{CVA6-based PoC.} CVA6 is an open-source RISC-V CPU initially designed as an Application class processor targeting Linux-capable systems \cite{zaruba2019}. Over time, CVA6 was extended with RISC-V Hypervisor support \cite{Sa2023, shaheen} and more recently, the microarchitecture was modified to allow the seamless removal of the MMU and related microarchitectural logic (e.g., TLB, PTW, etc.). This last addition allows the instantiation of CVA6 implementations in an MCU-like style, endowing the Supervisor mode with a PMP (i.e., SPMP) instead of an MMU. On top of this, we have implemented the SPMP for the hypervisor extension, which we will make open-source soon. We plan to extend this version with WorldGuard support and implement the extensions proposed in this paper. We argue we will provide the first open-source reference CPU/MCU featuring these state-of-the-art technologies. 

\mypara{Evaluation.} After implementing the CVA-based PoC, we aim at assessing and evaluating the system for multiple metrics and study different trade-offs. We highlight context- / VM-switch time, interrupt latency, inter-world/VM interference, and hardware costs among these target metrics. We plan to complement microbenchmark experiments with application benchmarks. We plan to assess and study different trade-offs and system configurations, such as the one reported in Section III.B, in the hopes of identifying the sweet spot in term of cost-function quandary.

\mypara{Resource Side Protection.} While in this paper we focus exclusively on CPU initiator-side protection (namely PMP and WID-control structures), WID-based checks are either performed in interconnect or resource-local. In this case, (i) introduced latencies, (2) cost-function, (3) and configuration- and security-model are key aspects that are highly influenced by architecture-/design of the checking structures, as the different models used currently in IOPMP already highlighted \cite{riscv-iopmp-spec}. We would like to emphasize that the establishment of a reference programming model for the checkers notably decides on ease-of-use, and consequently adoption at scale, of a WorldGuard-based isolation solution. We reserve these topics for future work.



\section{Conclusion}

RISC-V has been advocated as a game-changer ISA for next-generation automotive computing systems. In this paper, we provided our critical perspective on why RISC-V is not yet well-suited to address the requirements of next-generation automotive Zone and Domain controllers, particularly in light of the ISO21434 directives. Based on our perspective, we proposed a first set of ISA extensions, at the CPU initiator-side, to address such limitations and discussed the planned roadmap to implement a full open-source hardware-software proof-of-concept (PoC). Nevertheless, we argue this is just a first step towards aiming at fully defining an automotive-ready reference RISC-V real-time MCU computing platform.

\bibliographystyle{IEEEtran}

\end{document}